\documentclass[]{interact}

\usepackage{epstopdf}% To incorporate .eps illustrations using PDFLaTeX, etc.
\usepackage{subfigure}% Support for small, `sub' figures and tables

\usepackage{natbib}% Citation support using natbib.sty
\bibpunct[, ]{(}{)}{;}{a}{}{,}% Citation support using natbib.sty
\renewcommand\bibfont{\fontsize{10}{12}\selectfont}% Bibliography support using natbib.sty

\theoremstyle{plain}% Theorem-like structures provided by amsthm.sty

\theoremstyle{definition}

\theoremstyle{remark}

\usepackage{hyperref}
\usepackage{algorithm}
\usepackage{algpseudocode}

\def\mb{\mathbb}

\DeclareMathOperator*{\argmin}{arg\,min}
\DeclareMathOperator*{\argmax}{arg\,max}

\def\mb{\mathbb}

\newcommand{\be}{\begin{equs}}
\newcommand{\ee}{\end{equs}}

\begin{document}

\articletype{ARTICLE TEMPLATE}

\title{Bayesian Spike Train Inference via Non-Local  Priors}

\author{
\name{Abhisek Chakraborty\thanks{Author Email: cabhisek@stat.tamu.edu}}
\affil{Department of Statistics, Texas A\&M University, College Station, TX 77843, USA}
}

\maketitle

\begin{abstract}
Advances in neuroscience have enabled researchers to measure the activities of large numbers of neurons simultaneously in behaving animals. We have access to the fluorescence of each of the neurons which provides a first-order approximation of the neural activity over time. Determining a neuron’s exact spike from this fluorescence trace constitutes an active area of research within the field of computational neuroscience. We propose a novel Bayesian approach based on a mixture of half-non-local prior densities and point masses for this task. Instead of a computationally expensive MCMC algorithm, we adopt a stochastic search-based approach that is capable of taking advantage of modern computing environments often equipped with multiple processors, to explore all possible arrangements of spikes and lack thereof in an observed spike train. It then reports the highest posterior probability arrangement of spikes and posterior probability for a spike at each location of the spike train. Our proposals lead to substantial improvements over existing proposals based on $L_1$ regularization, and enjoy comparable estimation accuracy to the state-of-the-art $L_0$ proposal, in simulations, and on recent calcium imaging data sets. Notably, contrary to optimization-based frequentist approaches, our methodology yields automatic uncertainty quantification associated with the spike-train inference.
\end{abstract}


\section{Introduction}

When a neuron fires, calcium floods the cell. To measure this influx of calcium inside the cells, calcium imaging techniques make use of 
fluorescent calcium indicator molecules (\cite{Theis}, \cite{Rupprecht}). Thus, a neuron's calcium fluorescence trace can be seen as a first-order approximation of its activity level over time. However, typically fluorescence trace is not particularly interesting on its own. Instead, it is used to determine the underlying activity level, that is, the specific times at which the neuron spiked. Inference of spike times from fluorescence traces amounts to a challenging deconvolution problem, and research in computational neuroscience has been focused on solving it (  \cite{Vogelstein2009}; \cite{Vogelstein2010}; \cite{Pnevmatikakis}; \cite{NIPS2016_fc2c7c47}; \cite{Theis}; \cite{NIPS2016_fc2c7c47}; \cite{Friedrich2017}; \cite{Jewell2018}, \cite{Rupprecht}. 

% Model? 
We will work with an auto-regressive model for calcium dynamics that has been taken into consideration by several authors in the recent literature (\cite{Vogelstein2010}; \cite{Pnevmatikakis}; \cite{NIPS2016_fc2c7c47}; \cite{Friedrich2017}, \cite{Jewell2018}). We closely follow the notations introduced in \cite{Jewell2018}. Our model assumes that $y_t$, the fluorescence at the $t$-th time-step, is a noisy realization of $c_t$, the unobserved underlying calcium concentration at the $t$-th time-step. A $p$-th-order auto-regressive process governs the calcium concentration decay in the absence of a spike at the $t$-th time-step ($s_t$ = 0). The calcium concentration rises, nevertheless, if a spike occurs at the $t$-th time-step ($s_t>0$). Thus,
\begin{align}\label{eqn:ar_p}
    & y_t = \beta_0 + \beta_1 c_t + \varepsilon_t, \quad \varepsilon_t\stackrel{ind}{\sim} \mbox{N}(0, \sigma^2),\quad t = 1,\ldots,T;\notag\\
    & c_t = \sum_{i=1}^p \gamma_{i} c_{t-i} + s_t,\quad t = p+1,\ldots, T ;
\end{align}
In \ref{eqn:ar_p}, the quantities $\gamma_1,\ldots,\gamma_p$  are the parameters in the auto-regressive model. Note that the quantity $y_t$ in \ref{eqn:ar_p} is observed; all other quantities are unobserved. Since we would like to know whether a spike occurred at the $t$-th time-step, the parameter of interest is $s_t=c_t - \sum_{i=1}^p \gamma_{i} c_{t-i}\geq 0$.  For ease of exposition, following common practice in the literature \cite{Jewell2018}, we assume $\beta_0 = 0$ and $\beta_1 = 1$ in \ref{eqn:ar_p}. This assumption is made without loss of generality, since $\beta_0$ and $\beta_1$ can be estimated from the data, and the observed fluorescence $y_1,\ldots, y_T$ centered and scaled accordingly. Further, primarily we focus on the AR(1) case, where $\gamma_1= \gamma\in(0,1)$ and $\gamma_i = 0\ \text{for}\ i\geq 2$. Notably, calcium imaging data sets are usually in high dimensions where the spike train is observed for few thousand time-points. Spikes, however, only show up at a select few time points. In other words, the majority of the entries in $(s_1, \ldots, s_T)^{\prime}$ are zero. This is the sparsity assumption that we need to enforce, and consequently the major goal is to locate the time-points with $s_t=c_t -\gamma c_{t-1}>0$, equivalently, the time-points with a spike. 

%Penalised likelihood methods?
Many popular penalized likelihood techniques that were first developed for linear regression have also been used to spike train inference. \cite{Vogelstein2010} proposed to impose the LASSO penalty (\cite{LASSO}) on $s_2, \ldots, s_T$. \cite{NIPS2016_fc2c7c47} and \cite{Friedrich2017} instead
consider a closely-related problem that results from including an additional LASSO penalty for the initial calcium concentration $c_1$. More recently, \cite{Jewell2018} proposed to put the $L_0$ penalty on  $s_2, \ldots, s_T$, and developed an algorithm to obtain the exact global solution for the resulting optimization problem. Further, to ensure computation ease, they solve the optimization problem after removing the positivity constraint $s_t\geq0$. In reality, there isn't much of a difference between including and disregarding the positivity requirement because the solutions for real-world applications typically are the same for appropriate decay rate $\gamma$. This cutting-edge method \cite{Jewell2018} guarantees significant advantages in the spike detection task over its $L 1$ regularization-based rivals.

%Bayesian methods? Problem with Bayesian methods.
A few Bayesian solutions have also been put out to deal with this issue. \cite{Vogelstein2010} proposed a  non-negative deconvolution filter to infer the approximately most likely spike train of each neuron, given the fluorescence observations.  \cite{Pnevmatikakis2013} proposed discrete time algorithms to find the existence of a spike at each time bin using Gibbs methods, as well as continuous time algorithms that sample over the number of spikes and their
locations at an arbitrary resolution using Metropolis-Hastings
methods for point processes. \cite{NIPS2016_fc2c7c47} suggested spike inference from calcium imaging  data using a sequential monte carlo method (\cite{Moral}).  For a more detailed review, we refer to the recent thesis (\cite{Adams16}). Notably,  Bayesian methods are  less utilized than their frequentist counterparts, as the computational load of the associated Monte Carlo Markov Chain (MCMC) procedures often renders these approaches infeasible. This leads to the question of whether it's possible to develop a reasonably fast Bayesian methodology that can outperform dominant frequentist algorithms.

%Our contribution?
In this paper, we propose a new approach for this task based on a novel application of  a mixture prior on spike variables that has  a point mass at zero and a half  non local density (\cite{Johnson2010}), in the auto-regressive model for calcium dynamics. These non local priors, although introduced in the context of Bayesian hypotheses testing, have proven useful in varied setups, e.g, high-dimensional linear regression (\cite{Johnson2012}; \cite{Shin2018}), high-dimensional logistic regression (\cite{Amir2016}), survival analysis (\cite{Nikooienejad2020}), sequential inference (\cite{PRAMANIK2021102505}) etc. Here, we impose  inverse moment prior or product exponential moment prior on $s_t, t=2,\ldots, T$, and report the arrangements of spikes and lack thereof with highest posterior probability. The computationally burdensome MCMC process
is avoided by adapting a stochastic search based method (\cite{Shin2018}). 
%A general algorithm based on (\citeauthor{Amir2016}) is also provided to set the tuning parameter of the non local prior.

This article is structured as follows. In section 2, we introduce methodology based on a novel application of non-local prior densities. In section 3, we present elaborate and realistic simulation studies to compare the proposed methodology with the existing and state-of-the-art
penalised likelihood methods. Section 4 presents the spike train inference of calcium imaging data sets. Finally, we conclude.

\section{Methods}
\subsection{Model and Prior Specification}
Suppose we observe a spike train of calcium concentrations $\mathbf{y} =\{y_i\}_{t=1}^T$ indexed by time $t$. We primarily work with the AR(1) model for calcium dynamics:
\begin{align}\label{eqn:ar_1}
    & y_t = \beta_0 + \beta_1 c_t + \varepsilon_t, \quad \varepsilon_t\stackrel{ind}{\sim} \mbox{N}(0, \sigma^2),\quad t = 1,\ldots,T;\notag\\
    & c_t = \gamma c_{t-1} + s_t,\quad t = 2,\ldots, T ;
\end{align}
where we assume that $\beta_0 = 0,\ \beta_1 = 1$ without loss of generality. We formally define an \emph{arrangement} of the spikes and lack thereof  by $\mathbf{k} = \{k_1,k_2,\ldots, k_j\}$ where $2\leq k_1<k_2\ldots<k_j\leq T$ such that $s_t > 0$ if $t\in\mathbf{k}$ and $0$ otherwise. Next, we define a Bayesian hierarchical model in which $ \pi(\mathbf{y}\mid s_{(\mathbf{k})},\ c_{1:T})$  is the likelihood of $(s_{(\mathbf{k})}, c_{1:T})$ obtained from the model in \ref{eqn:ar_1}, $\pi_{\mathbf{k}}(\cdot)$ is the prior on the spike variables $s_{(\mathbf{k})}$,  $\pi_c(\cdot)$ is a  compactly supported flat prior on $c_{1:T}$, and finally $p(\cdot)$ is the prior for arrangement $\mathbf{k}$. The posterior probability for arrangement $\mathbf{j}$ according to the Bayes rule is expressed as
\begin{align*}
    p(\mathbf{j}\mid \mathbf{y}) = \frac{p(\mathbf{j})\ m_{\mathbf{j}}(\mathbf{y})}{\sum_{k\in \mathcal{J}}p(\mathbf{k})\ m_{\mathbf{k}}(\mathbf{y})}
\end{align*}
where $\mathcal{J}$ is the set of all possible arrangements and the marginal probability of the data under the arrangement $\mathbf{j}$ is computed by
\begin{align*}
m_{\mathbf{j}}(\mathbf{y}) = \int \pi(\mathbf{y}\mid s_{(\mathbf{j})})\ \pi_{\mathbf{j}}(s_{(\mathbf{j})})\ \pi_{\mathbf{c}}(c_{1:T}) ds_{(\mathbf{j})} d\mathbf{c}.    
\end{align*}
Notably, the prior density for $\pi_{\mathbf{k}}(\cdot)$ and the prior  on the arrangement space $p(\mathbf{j})$ impact the overall performance of the spike detection mechanism and the number of spikes in the resulting arrangement, and therefore we exercise ardent care in our choices of these priors.

First, we discuss the prior $p(\cdot)$ on the spike arrangement space. Note that, the events $\{t\in \mathbf{k}\},\ t = 2,\ldots, T$ can be modeled via independent Bernoulli trials. In particular, if we assume $\mbox{P}(t\in \mathbf{k}) = \theta\in(0, 1)$, then the resulting marginal probability for the arrangement $\mathbf{k}$ becomes $\int_{0}^1 \theta^k (1-\theta)^{T-1-k}\ \pi(\theta) d\theta$, where $\pi(\theta)$ is the prior on $\theta$. A completely conjugate choice of $\pi(\theta)$ would be $\beta(a, b)$, and that yields 
\begin{align}
    p(\mathbf{k}) = \int_{0}^1 \theta^k (1-\theta)^{T-1-k}\ \pi(\theta) d\theta = \frac{\beta(a + k ,\ b + T-1-k)}{\beta(a, b)}.
\end{align}
For most practical purposes, we suggest to choose $a=1$ and $b = 1$ which ensures that we do not necessarily assign  low prior probability to arrangements with many spikes. 

Next, we focus on the prior specification on the spike variables $s_t,\ t = 2, \ldots, T$. Here we put two-component mixture that has a spike at zero and a half-non-local density. In particular, we consider either of the two classes of half-non-local prior densities on $\mb{R}^{+}$, obtained via restricting the non-local priors (\cite{Johnson2010}, \cite{Johnson2012}, \cite{Shin2018}) with full support on $\mb{R}$, as follows:
\begin{align}\label{prior_nonlocal}
    &\pi(s) \ \propto \ s^{-(r+1)}\exp{\bigg(-\frac{\tau}{s^2}\bigg)}, \ s>0,\quad [\text{Inverse moment prior (iMOM)}];\\
    & \pi(s) \ \propto \ s^{-(r+1)}\exp{\bigg(-\frac{s^2}{2\sigma^2\tau} - \frac{\tau}{s^2}\bigg )}, \ s>0,\ [\text{Exponential moment prior (eMOM)}].
\end{align}
Here the $(\tau, r)$ are  prior hyper-parameters. While the hyper parameter $r$ is comparable to the shape parameter in the inverse gamma distribution and controls the tail behavior of the density, the hyper parameter $\tau$ represents a scale parameter that affects the prior's dispersion around 0. Notably, large values of the spikes are not penalized by this prior, in contrast to the majority of penalized likelihood approaches. Consequently, it does not always impose harsh penalties on arrangements that have a lot of spikes. In general, we found that $r = 1$ and $\tau = 0.25$ are good default values. %\textcolor{blue}{Include some pictures of the prior. May want to include a discussion on hyper-parameter tuning} 
Finally, note that the model in \ref{eqn:ar_1}
 can be completely expressed by means  of the minimal set of parameters parameters $c_1,\ \sigma^2, \ s_{2:T}$. We complete our prior specification by putting a compactly supported  non-informative prior on $c_1$ of the form 
 \begin{align}\label{prior_c}
     \pi_{\mathbf(c)}(c_1)\sim \ \mbox{Unifrom}(0, c_{max})\quad \text{where}\ c_{max}\ \text{is large enough},
 \end{align}
and 
\begin{align}\label{prior_sigma2}
    \pi(\sigma^2) \equiv \mbox{Inverse-Gamma}(a_0,\ b_0).
\end{align}
So, the equations \eqref{eqn:ar_1}-\eqref{prior_sigma2} complete our model and prior specification for the studying calcium dynamics.

\subsection{Posterior Inference}
Since we monitor the spike train over a huge number of time points, full posterior sampling with the current MCMC techniques is very costly and frequently not practicable. To overcome this problem, based on \cite{Shin2018}, we suggest a scalable stochastic search strategy for quickly locating areas of high posterior probability and determining  maximum a posteriori arrangement of spikes and absence of spikes. To that end, first note that $\hat{\mathbf{k}}$ denote the maximum a posteriori arrangement:
\begin{align}\label{map}
    \hat{\mathbf{k}} = \argmax_{\mathbf{k}\in\Gamma} \pi(\mathbf{k}\mid\mathbf{y}),
\end{align}
where $\Gamma$ is the set of all arrangements assigned non-zero prior probability. However,  we cannot ensure that a local maxima found by a deterministic search algorithm is the global maximum until we search all feasible arrangements in $\Gamma$ because the optimization problem in \eqref{map} is NP-hard.

\subsubsection*{Stochastic Search Algorithms}
In the context of model selection tasks, to overcome this problem, \cite{S3} proposed a shotgun stochastic search  algorithm in an attempt to identify the global maximum with high probability. This approach defines a neighbourhoods of the form $\mbox{N}(\mathbf{k})= \{\Gamma^{-},\ \Gamma^0,\ \Gamma^{+}\}$ where $\Gamma^{-} = \{\mathbf{k}\setminus j:\ j\in\mathbf{k}\}$, $\Gamma^{+} = \{\mathbf{k}\cup j: j\in\mathbf{k}^c\}$ and $\Gamma^{0} = \{(\mathbf{k}\setminus j)\cup l: j\in\mathbf{k},\ l\in \mathbf{k}^c\}$, and proposed to the stochastic search \textbf{Algorithm} \ref{alg:s3}.

\begin{algorithm}
\caption{Shotgun Stochastic Search (SSS), \cite{S3}}\label{alg:s3}
%\begin{algorithmic}
$\mathbf{(1)}$ Choose an $\mathbf{k}^{(1)}$ and total number of iterations $N$.\\
$\mathbf{(2)}$ For $i\in\{1, \ldots, N-1\}$\\
----- (a) calculate $\pi(\mathbf{k}\mid\mathbf{y})$ for all $\mathbf{k}\in \mbox{N}(\mathbf{k}^{(i)})$,\\
----- (b) select $\mathbf{k}^{-}$, $\mathbf{k}^{0}$, and $\mathbf{k}^{+}$ from $\Gamma^{-},\ \Gamma^0,\ \Gamma^{+}$ with probabilities $\pi(\cdot\mid\mathbf{y})$,\\
----- (c) select $\mathbf{k}$ from $\{\mathbf{k}^{-}$,\ $\mathbf{k}^{0},\ \mathbf{k}^{+}\}$ with probabilities $\{\pi(\mathbf{k}^{-}\mid\mathbf{y}),\ \pi(\mathbf{k}^{0}\mid\mathbf{y}),\ \pi(\mathbf{k}^{+}\mid\mathbf{y})\}$.\\
$\mathbf{(3)}$  Choose the arrangement with largest posterior probability among $\{\mathbf{k}^{(i)},\ i=1,\ldots, N\}$. 

%\end{algorithmic}
\end{algorithm}

Shotgun Stochastic Search is efficient in locating regions of high posterior probability in the arrangement space, but because it necessitates the assessment of marginal probabilities for arrangements in $\Gamma^{-},\ \Gamma^0,\ \Gamma^{+}$ in each iteration, its computational cost is still significant. The largest computational expense is due to the evaluation of
marginal likelihood for arrangements in $\Gamma^0$ since $|\Gamma^0|= |\mathbf{k}|(T - |\mathbf{k}|)$. In an attempt to remedy this issue, in the context of model selection exercises, \cite{Shin2018} proposed a simplified Shotgun Stochastic Search that only considers arrangements in $\Gamma^{-},\ \Gamma^{+}$  which have cardinality $(T - |\mathbf{k}|)$ and $|\mathbf{k}|$, respectively. However, by ignoring $\Gamma^0$ in the sampling updates, we increase the likelihood that the algorithm will become trapped in a local maxima and make it less likely to explore the intriguing regions with high posterior  probability. To solve this issue, a temperature parameter that is similar to simulated annealing was included, enabling the algorithm to explore a wider range of configurations.
 
Ignoring arrangements in $\Gamma^0$ undoubtedly reduces the computational burden of the Shotgun Stochastic Search algorithm. But when $T$ is very large, computing the posterior probability for each iteration is still computationally expensive. To further alleviate this computational challenge, following \cite{Shin2018}, we  adapt ideas from Iterative Sure Independence Screening \cite{FanLV} and consider only those
time-points which have a large correlation with the residuals of the current arrangement. To make it precise, first note that the model in \ref{eqn:ar_1} can be expressed as
\begin{align*}
    \mathbf{y}= \Omega\ (c_1, s_2, \ldots s_T)^{\prime} + \mathbf{\varepsilon}
\end{align*}
where $\mathbf{y} = (y_1, y_2,\ldots, y_T)^{\prime}$, $\mathbf{\varepsilon} = (\varepsilon_1, \ldots, \varepsilon_{T})^{\prime}$ and  $\Omega = (\Omega_1,\ \Omega_2,\ \ldots,\ \Omega_T)^{\prime}$ is a $T\times T$ matrix of the form:
\begin{align*}
    \begin{bmatrix}
    1 & 0 & 0 & \ldots &0\\
    \gamma & 1 & 0 & \ldots &0\\
    \gamma^2 & \gamma & 1 & \ldots &0\\
    \gamma^3 & \gamma^2 & \gamma & \ldots &0\\
    \ldots & \ldots &\ldots &\ldots &\ldots\\
    \gamma^{T-1} & \gamma^{T-2}& \gamma^{T-3} & \ldots &1\\
    \end{bmatrix}
\end{align*}
where $\gamma\in(0,1)$. We compute the quantities $r_{\mathbf{k}}^{\prime} \Omega_j$, where $r_{\mathbf{k}}$ is the residual of arrangement $\mathbf{k}$, for $j = 1, \ldots, T$, after
iteration $j$ of the modified algorithm, and then restrict attention to time-points for which $|r_{\mathbf{k}}^{\prime} \Omega_j|$ is large (we assume that the columns of $\Gamma$ have been standardized). This yields a scalable algorithm even when the number of time points $T$ is large. 

The resulting Simplified Shotgun Stochastic Search with Screening (S5) algorithm is described in \textbf{Algorithm} \ref{alg:s5}. In this algorithm, $S_{\mathbf{k}}$ is the union of time-points in $\mathbf{k}$ and the top $M$ time-points
obtained by screening using the residuals from arrangement $\mathbf{k}$. So, the screened neighborhood of arrangement $\mathbf{k}$ can be defined as $\mbox{N}_{\rm screen}(\mathbf{k}) =\{ \Gamma_{\rm screen}^{+},\ \Gamma^{-}\}$, where $\Gamma_{\rm screen}^{+} = \{\mathbf{k}\cup\{j\}: j\in\mathbf{k}^c\cap S_{\mathbf{k}}\}$. The adapted S5 algorithm only considers the addition of $M$ new time-points in each iteration, which dramatically reduces the computational burden to construct $\Gamma_{\rm screen}^{+}$. In order to prevent re-evaluation and further improve computing performance, we keep the posterior probabilities of visited arrangemnts.

\begin{algorithm}
\caption{Simplified shotgun stochastic search algorithm with screening (S5), adapted from  \cite{Shin2018}}\label{alg:s5}
%\begin{algorithmic}
$\mathbf{(1)}$ Set a temperature schedule $t_1 > t_2 > . . . > t_L > 0$.\\
$\mathbf{(2)}$ Choose an $\mathbf{k}^{(1, 1)}$, a set of time-points after screening $\mathbf{S}_\mathbf{k}^{(1, 1)}$ based on $\mathbf{k}^{(1, 1)}$ and total number of iterations $N$.\\
$\mathbf{(3)}$ For $l\in\{1, \ldots, L\}$,\\
----- For $i\in\{1, \ldots, J-1\}$,\\
----- ----- (a) calculate $\pi(\mathbf{k}\mid\mathbf{y})$ for all $\mathbf{k}\in \mbox{N}_{\rm screen}(\mathbf{k}^{(1, 1)})$,\\
----- ----- (b) select $\mathbf{k}^{-}$, and $\mathbf{k}^{+}$ from $\Gamma^{-}, \ \Gamma^{+}_{\rm screen}$ with probabilities $\pi(\cdot\mid\mathbf{y})^{1/t_l}$,\\
----- ----- (c) select $\mathbf{k}^{i+1, l}$ from $\{\mathbf{k}^{-}$,\ $\ \mathbf{k}^{+}\}$ with probabilities $\{\pi(\mathbf{k}^{-}\mid\mathbf{y})^{1/t_l},\  \pi(\mathbf{k}^{+}\mid\mathbf{y})^{1/t_l}\}$.\\
----- ----- (d) update the set of variables considered to $\mathbf{S}_\mathbf{k}^{(i+1, l)}$ to be union of time-points in $\mathbf{k}^{i+1, l}$ and top $M_n$ time-points according to $\{|r_{\mathbf{k}^{i+1,l}}^{\prime} \Omega_j|, j = 2,\ldots, T\}$.

%\end{algorithmic}
\end{algorithm}

\subsubsection*{Calculation of the Marginal Likelihood}
Closed form expressions for posterior arrangement probabilities based on modified peMoM priors and modified piMoM priors are not available. We estimate the posterior arrangement  probabilities
using Laplace approximations, given by
\begin{align}
    m_{\mathbf{k}}(\mathbf{y}) = (2\pi)^{|\mathbf{k}|/2}\ \pi(\mathbf{y}\mid \hat{s}_{\mathbf{k}})\ \pi(\hat{s}_{\mathbf{k}})\ |G(\hat{s}_{\mathbf{k}})|^{-1/2},
\end{align}
where $\hat{s}_{\mathbf{k}}$ is MAP of ${s}_{\mathbf{k}}$, $ G(\hat{s}_{\mathbf{k}})$ is the Hessian of the negative log posterior density
\begin{align*}
    g( {s}_{\mathbf{k}}) = -\log\pi(\mathbf{y}\mid {s}_{\mathbf{k}}) - \log\pi( {s}_{\mathbf{k}}),
\end{align*}
evaluated at $\hat{s}_{\mathbf{k}}$.  Following \cite{Shin2018}, we use the limited memory version of the Broyden–Fletcher–Goldfarb–Shanno optimization algorithm (L-BFGS) (\cite{Liu1989OnTL}) to find the MAP.

With this Laplace approximation of the arrangements probabilities, we utilize  the stochastic algorithm discussed earlier to search the arrangement space to complete our posterior analysis. We report two summaries of our posterior analysis - (i) the highest posterior probability  arrangement  which is defined as the arrangement having the highest posterior probability among all visited arrangements, and (ii) the posterior  probability for a spike at time point $t$ which is defined as the sum of posterior probabilities of all arrangements that have a spike at time $t$.
%\textcolor{red}{Check!} 
Since we have the complete model prior specification, and corresponding computational framework for posterior analysis in place, we are in a position to investigate the empirical performance of the proposed methodology.

\subsection{Choice of the decay parameter $\gamma$ in \eqref{eqn:ar_1}}\label{ssec:decay_parameter}
In literature, there are two common methods for estimating the decay parameter $\gamma\in(0, 1)$ in \eqref{eqn:ar_1}, that controls the rate of exponential decay of the calcium concentration. (i) Pnevmatikakis et al. (2013), Friedrich and Paninski (2016), and Friedrich, Zhou and Paninski (2017) propose to select the exponential decay parameter $\gamma$ based on the auto-covariance function. We
refer the reader to Friedrich, Zhou and Paninski (2017) and Pnevmatikakis et al. (2016) for additional details. (ii) Alternatively, Jewell and Witten (2018) advise choosing a segment $y_{a:b}$ manually first if it  demonstrates exponential decline upon visual inspection.. Next, they estimate $\gamma$ by
\begin{align*}
    \hat{\gamma} = \argmin_{\gamma\in(0, 1)} \bigg\{\min_{c_a, c_t = \gamma c_{t-1}, t = a+1, \ldots, b }\bigg\{ \frac{1}{2}\sum_{t=a}^b (y_t - c_t)^2\bigg\}\bigg\}.
\end{align*}
Numerical optimization can be used to accomplish this.

\section{Simulation Studies}
In this section, we use simulated data to demonstrate the performance advantages of the proposed non local prior based methodologies over two competing frequentist approaches:

(i). The $L_1$ proposal in \cite{NIPS2016_fc2c7c47} and \cite{Friedrich2017}, which involves a single tuning parameter $\lambda$. These approaches solve the optimization problem:
\begin{align}\label{eqn:L1}
    \min_{c_1,\ldots, c_T, s_2,\ldots, s_T}\bigg\{\frac{1}{2}\sum_{t=1}^T (y_t - c_t)^2 + \lambda|c_1| +\lambda \sum_{t=2}^T|s_t| \bigg\} \quad \text{subject to}\quad s_t = c_t-\gamma c_{t-1}\geq 0,
\end{align}
where the hyper-parameter $\lambda$ is fixed via cross-validation.

(ii). The state-of-the-art  $L_0$ proposal in \cite{Jewell2018}, which too involves a single tuning parameter $\lambda$, that posses the following optimization problem:
\begin{align}\label{eqn:L0}
    \min_{c_1,\ldots, c_T, s_2,\ldots, s_T}\bigg\{\frac{1}{2}\sum_{t=1}^T (y_t - c_t)^2\ + \ \lambda \sum_{t=2}^T \delta(s_t \neq 0) \bigg\},
\end{align}
where $\delta(\cdot)$ is the Dirac delta measure, and the hyper-parameter $\lambda$ is again fixed via cross-validation.

We measure performance in spike detection of each method based on several criteria: (i) accuracy, (ii) sensitivity,  (iii) specificity, and (iv) false discovery rate. To that end,  suppose $c_1,c_2,\ldots, c_T$ denote the true calcium concentration observed over time $t=1, 2, \ldots, T$, and we define $s_t = c_t - \rho c_{t-1}, t=2, \ldots, T$. Then, we can define
{\footnotesize
\begin{align}
&\text{Accuracy} = \frac{1}{T-1}\bigg[\sum_{t=2}^T\bigg\{ \delta(s_t > 0,\ \hat{s}_t >0) + \delta(s_t = 0,\ \hat{s}_t =0)\bigg\}\bigg]\times 100 \% \ ;\notag\\
&\text{Sensitivity} = \frac{\sum_{t=2}^T \delta(s_t > 0,\ \hat{s}_t >0)}{\sum_{t=2}^T\big[ \delta(s_t > 0,\ \hat{s}_t >0) + \delta(s_t > 0,\ \hat{s}_t =0)\big]}\times 100 \% \ 
= \frac{\sum_{t=2}^T  \delta(s_t > 0,\ \hat{s}_t >0)}{\sum_{t=2}^T\big[ \delta(s_t > 0)\big]}\times 100 \% \ ;\notag\\
&\text{Specificity} 
= \frac{\sum_{t=2}^T \delta(s_t = 0,\ \hat{s}_t =0)}{\sum_{t=2}^T\big[ \delta(s_t = 0,\ \hat{s}_t =0) + \delta(s_t > 0,\ \hat{s}_t =0)\big]}\times 100 \%\  
= \frac{\sum_{t=2}^T \delta(s_t = 0,\ \hat{s}_t =0)}{\sum_{t=2}^T\big[ \delta(\hat{s}_t =0) \big]}\times 100 \% \ ;  \notag\\ 
&\text{False discovery rate (FDR)} 
= \frac{\sum_{t=2}^T \delta(s_t = 0,\ \hat{s}_t >0)}{\sum_{t=2}^T\big[ \delta(\hat{s}_t >0) \big]}\times 100 \% \ ; 
\end{align}
}
where $\delta(\cdot)$ is the Dirac delta measure, and we favour larger values of true spike detection rate and smaller values of false spike detection rate. Further, unlike the competing frequentist penalization methods, our proposed non local prior based approaches also yield (i) posterior probability  associated with the maximum a posteriori arrangement of spikes and lack thereof, and (ii) posterior probability for a spike at each location of the spike train.

\begin{table}[!htb]
\caption{Accuracy, sensitivity, and specificity for different methods averaged over $50$ simulated data sets for $\gamma = 0.96$, $\sigma=0.15$ and varying values of  $T\in\{ 2000, 2500\}$. }\label{tab:spike_detection96}
\small\setlength\tabcolsep{4.5pt}
  \begin{tabular}{ccccccccc}
    \hline
      \multicolumn{1}{c}{} &
      \multicolumn{4}{c}{$T = 2000$} &
      \multicolumn{4}{c}{$ T = 2500$}\\
    \hline
      \multicolumn{1}{c}{} &
      \multicolumn{1}{c}{Accuracy} &
      \multicolumn{1}{c}{Sensitivity} &
      \multicolumn{1}{c}{Specificity} &
      \multicolumn{1}{c}{FDR} &
     \multicolumn{1}{c}{Accuracy} &
      \multicolumn{1}{c}{Sensitivity} &
      \multicolumn{1}{c}{Specificity}&
      \multicolumn{1}{c}{FDR} \\
    \hline
    iMOM &99.97 & 97.93 &99.99 &00.00 &99.98 & 98.12 & 99.99 & 00.00\\
    eMOM &99.98&98.07 &100.00 &00.00 &99.98 &98.12&100.00&00.00\\
    \hline
    $L_0$ &99.98 &98.17 &99.99 &00.00 &99.98 &98.68&99.99&00.00\\
    \hline
    $L_1$ &97.18 &100.00 &97.15 &76.39 & 97.17&1.00&97.14&77.27\\
    \hline
  \end{tabular}
\end{table}

\begin{table}[!htb]
\caption{Accuracy, sensitivity, and specificity for different methods averaged over $50$ simulated data sets for $\gamma = 0.98$, $\sigma=0.15$ and varying values of  $T\in\{ 2000, 2500\}$. }\label{tab:spike_detection98}
\small\setlength\tabcolsep{4.5pt}
  \begin{tabular}{ccccccccc}
    \hline
      \multicolumn{1}{c}{} &
      \multicolumn{4}{c}{$T = 2000$} &
      \multicolumn{4}{c}{$ T = 2500$}\\
    \hline
      \multicolumn{1}{c}{} &
      \multicolumn{1}{c}{Accuracy} &
      \multicolumn{1}{c}{Sensitivity} &
      \multicolumn{1}{c}{Specificity} &
      \multicolumn{1}{c}{FDR} &
     \multicolumn{1}{c}{Accuracy} &
      \multicolumn{1}{c}{Sensitivity} &
      \multicolumn{1}{c}{Specificity}&
      \multicolumn{1}{c}{FDR} \\
    \hline
    iMOM &99.97 &97.20 &99.99 &00.00 & 99.98 &98.12 &99.99 &00.00\\
    eMOM & 99.97&97.20 &99.99 &00.00 &99.97 &98.12&99.99&00.00\\
    \hline
    $L_0$ &99.98 &98.10 &99.99 &00.00 &99.98 &98.53&99.99&00.00\\
    \hline
    $L_1$ &96.31 &100.00 &96.27 & 81.91& 96.35&100.00&96.31&79.67\\
    \hline
  \end{tabular}
\end{table}

Based on \cite{Jewell2018}, we generate $50$ simulated data sets according to \eqref{eqn:ar_1} with parameter settings $\gamma\in\{0.96, \ 0.98\}$, $T \in \{2000, \ 2500\}$, $\sigma = 0.15$, and $s_t \stackrel{i.i.d}{\sim} \mbox{Poisson}(0.01)$. On each simulated data set, we compare our methodology with the two competing methods with respect to the performance metric delineated earlier. The decay parameter $\gamma$ is estimated by the methods described in \ref{ssec:decay_parameter}. The $L_0$ (\ref{eqn:L0}) and $L_1$ (\ref{eqn:L1}) regularised approaches are utilized with a hyper parameter tuning step involving cross-validation. For the inverse moment prior and the exponential moment prior based approaches, we utilize the S5 algorithm with a Bernoulli-Uniform prior on the spike arrangement space. To ensure the stability of the stochastic search based algorithm, we utilize $10$ random starting points and a temperature schedule of $\{0.4, 0.5, \ldots, 1.0\}$. Also, the stochastic search was distributed in $50$ cores in order to expedite the calculations.

In Tables \ref{tab:spike_detection96} and \ref{tab:spike_detection98}, we present the performance metrics for the spike detection task averaged over $50$ simulated data sets, corresponding to our non-local prior based methods as well as the penalised likelihood based frequentest methods. All four methods perform extremely well in detecting  true spikes, as depicted by comparably high sensitivity and specificity figures across all simulation setups. However, the frequentest method based on $L_1$ regularization results in high number of false spike discovery, whereas the $L_0$ regularised method and the non-local prior based methods almost never discovered a false spike across the repeated simulations. This provides compelling empirical evidence that, our non-local prior based proposals lead to substantial improvements over the previous proposal based on $L_{1}$ regularization , and enjoys comparable estimation accuracy to the state-of-the-art $L_{0}$ proposal. Furthermore, contrary to the optimization based frequentest approaches, our methodologies yields automatic uncertainty quantification associated with the spike train inference. In particular, we provide the probability corresponding to the highest posterior probability arrangements of spikes and lack there of. Moreover, we provide the posterior probability of a spike at each time-point in the spike-train.

\begin{figure}[!htb]
    \centering
    {{\includegraphics[width=16cm, height = 6cm]{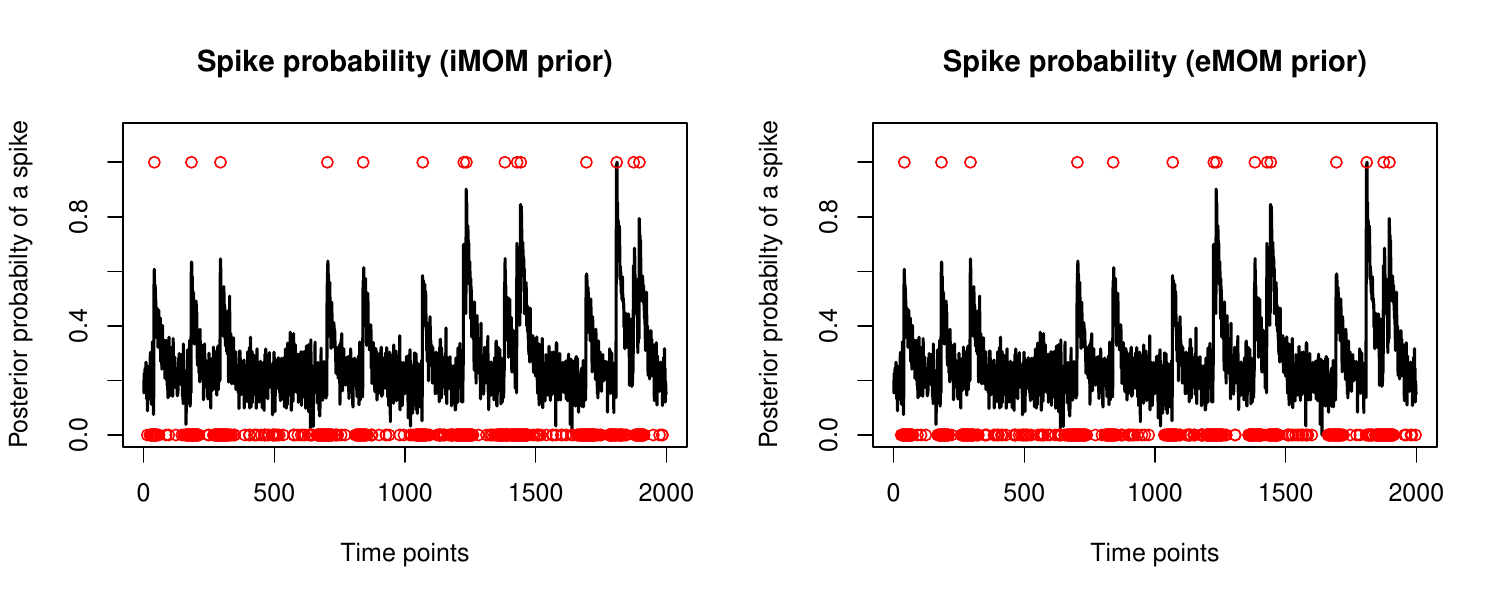} }}%
    \caption{\textbf{Simulated data:} \emph{One particular simulated data generated with specifications  $\gamma = 0.96,\ T = 2000, \ \sigma = 0.15$. The two panel displays the centered and scaled DF/F-transformed fluorescence re scaled to $(0,1)$ (in black lines) and the posterior probabilities of  spikes at different time points by red dots, for the iMOM, eMOM based methods respectively.}\label{sim_UQ}}
\end{figure}

\begin{figure}[!htb]
    \centering
    {{\includegraphics[width=16cm, height = 10cm]{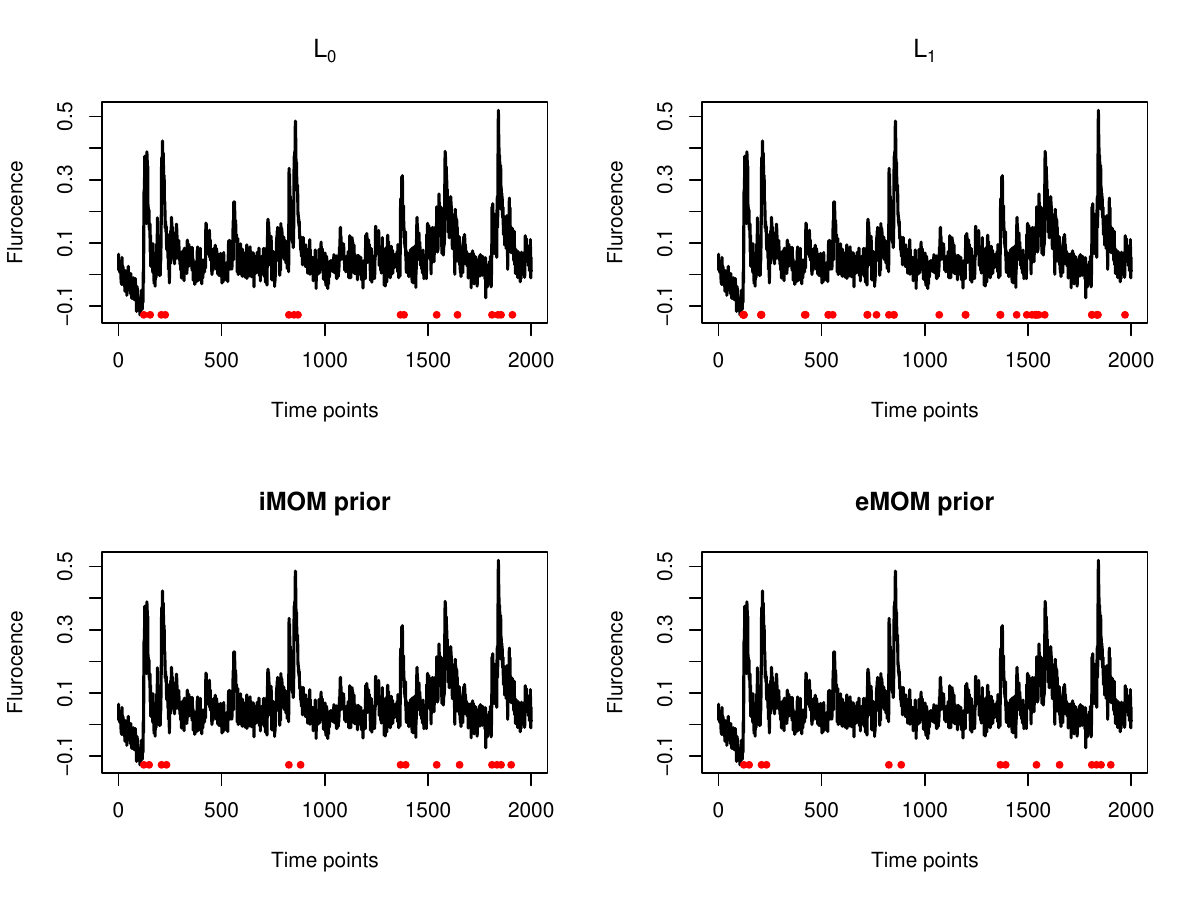} }}%
    \caption{\textbf{Mice data:} \emph{The calcium fluorescence data from mice in \cite{Theis}. Each panel displays the DF/F-transformed fluorescence ( in black lines ), the estimated spikes by the red dots. The four panels display results obtained via the $L_0$, $L_1$, iMOM, eMOM based methods respectively. }\label{data1_estimation}}
\end{figure}

In particular, for the sake of this presentation, we focus on a single simulated data set generated with $\gamma = 0.96,\ T = 2000, \ \sigma = 0.15$. For this data set, the $L_0$,  inverse moment prior, and exponential moment prior based approaches identified exactly same number of spikes at very similar and visually reasonable time points, presented as red dots in  figure \ref{data1_estimation}.  But, as observed in  much of previous literature (\cite{Jewell2018}), the $L_1$ regularised approach results in many, typically small in magnitude, false spike discoveries. Further, we observe that the  maximum a posteriori estimates corresponding to the inverse moment and exponential moment priors has posterior probabilities equal to $0.9999$ and $0.9997$ respectively. The posterior probability of a spike at the different time point corresponding to the inverse moment and exponential moment prior based approaches are presented as red dots in the two panels in figure \ref{sim_UQ}.

\begin{figure}[!htb]
    \centering
    {{\includegraphics[width=16cm, height = 10cm]{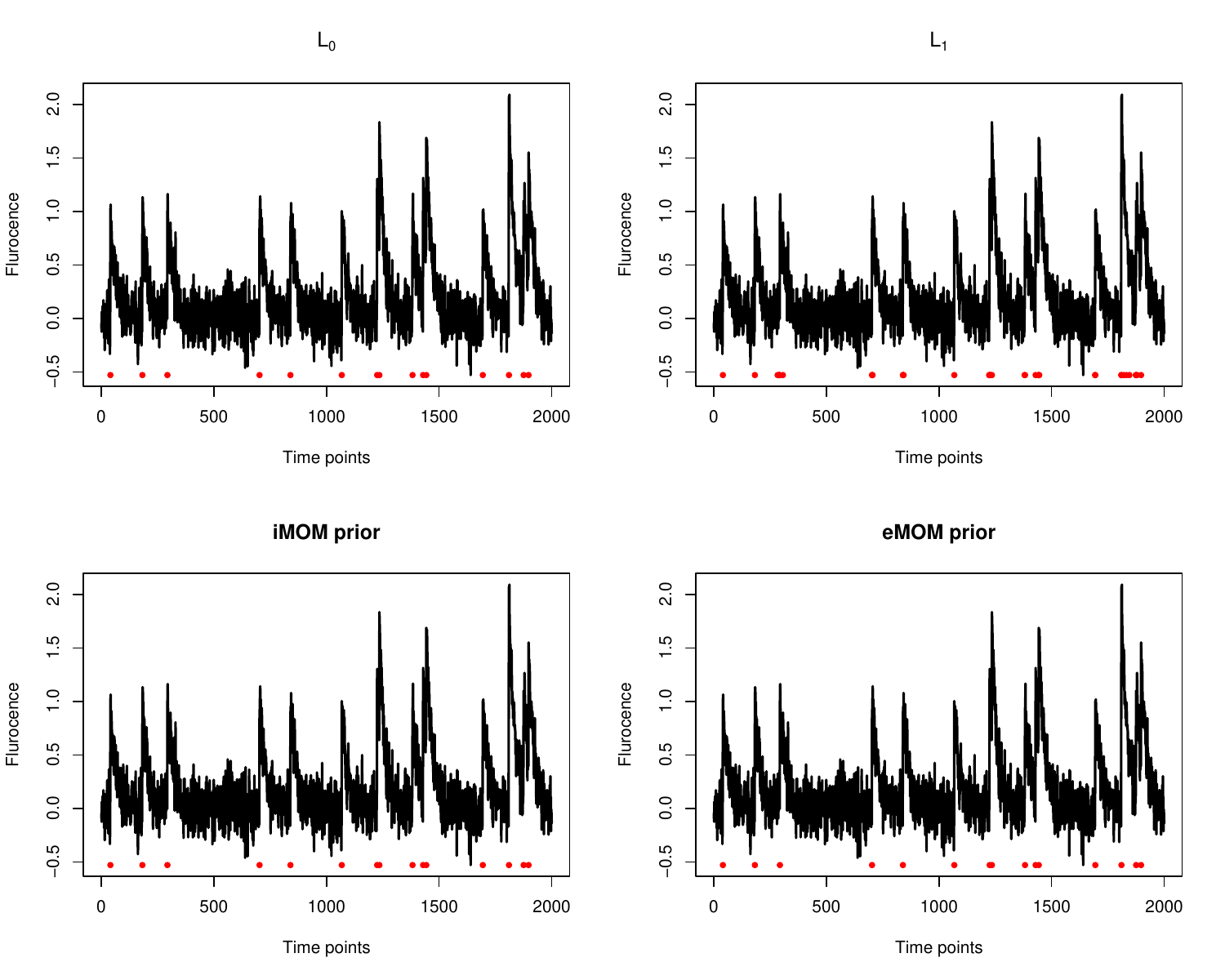} }}%
    \caption{\textbf{Simulated data:} \emph{One particular simulated data generated with specifications  $\gamma = 0.96,\ T = 2000, \ \sigma = 0.15$ . Each panel displays the DF/F-transformed fluorescence ( in black lines ), the estimated spikes by the red dots. The four panels display results obtained via the $L_0$, $L_1$, iMOM, eMOM based methods respectively. }\label{sim_estimation}}
\end{figure}

%\textcolor{blue}{Provide ( 1 + 1 + 2 + 2)  plots for one example data set, 
%S5 specifications, hyper-parameter specifications  and setups of S5 and others.
%}

%\clearpage
\section{Application to Calcium Imaging data sets}
Action potential derivation from neuronal calcium signals is complicated due to the lack of simultaneous measurements of action potentials and calcium signals. \cite{Rupprecht} compiled a huge and varied database (\href{https://github.com/HelmchenLabSoftware/Cascade/}{github.com/HelmchenLabSoftware/Cascade}) from publicly available and newly performed recordings in \emph{zebrafish} and \emph{mice}, consisting of 298 neurons' combined recordings over more than 35 hours, and encompassing a wide range of cell types, calcium markers, and signal-to-noise ratios. By performing simultaneous electrophysiological recordings and calcium imaging in both \emph{zebrafish} and \emph{mice}, \cite{Rupprecht} was able to broaden the range of data sets that were available. In \emph{zebrafish}, using the synthetic calcium indicators Oregon Green BAPTA-1 (OGB-1) and Cal-520 as well as the genetically encoded calcium indicator GCaMP6f, a total of 47 neurons in various telencephalic areas were recorded in the juxtacellular configuration in an explant preparation of the entire adult brain.. In head-fixed \emph{mice}, using the genetically encoded indication R-CaMP1.07, recordings were made in the hippocampus area CA3 while the subject was under anesthesia. 
Further, \cite{Rupprecht} examined freely available data sets and gathered information from raw movies. There were originally 157 neurons available, down from a total of 193 after thorough quality screening. In addition to their own recordings, \cite{Rupprecht} compiled 27 data sets with a combined total of 495,077 spikes and $\sim38$ hours of recording from 298 neurons, eight calcium indicators, and nine different brain areas in two species..

The data sets showed significant differences in recording times, imaging frame rates, and spike rates. Typical spike rates
spanned more than an order of magnitude, ranging from 0.4 Hz to
11.6 Hz, and frame rates varied between 7.7 Hz and $>160$ Hz . Using regularized deconvolution, \cite{Rupprecht} computed the
linear $\Delta F/F$ kernel evoked by the average spike and discovered that, even for data from the same calcium indicator, the area under the kernel curve varied significantly between data sets and was significantly smaller for data sets with inhibitory neurons. The issue faced by any algorithm that is supposed to analyze the data is shown by the fact that kernels displayed significant variance even among neurons within the same data set.

To demonstrate the performance the proposed  non local priors based approaches in spike estimation estimation and uncertainty quantification, we present the results for two particular data sets - one is a calcium fluorescence observed in mice \cite{Theis}, and the other for zebra fish (\cite{Rupprecht}). The $L_0$ (\ref{eqn:L0}) and $L_1$ (\ref{eqn:L1}) regularised approaches are utilized with a hyper parameter tuning step involving cross-validation. For the inverse moment prior and the exponential moment prior based approaches, we utilize the S5 algorithm with a Bernoulli-Uniform prior on the spike arrangement space. To ensure the stability of the stochastic search based algorithm, we utilize $10$ random starting points and a temperature schedule of $\{0.4, 0.5, \ldots, 1.0\}$. Also, the stochastic search was distributed in $50$ cores in order to expedite the calculations.

\begin{figure}[!htb]
    \centering
    {{\includegraphics[width=16cm, height = 6cm]{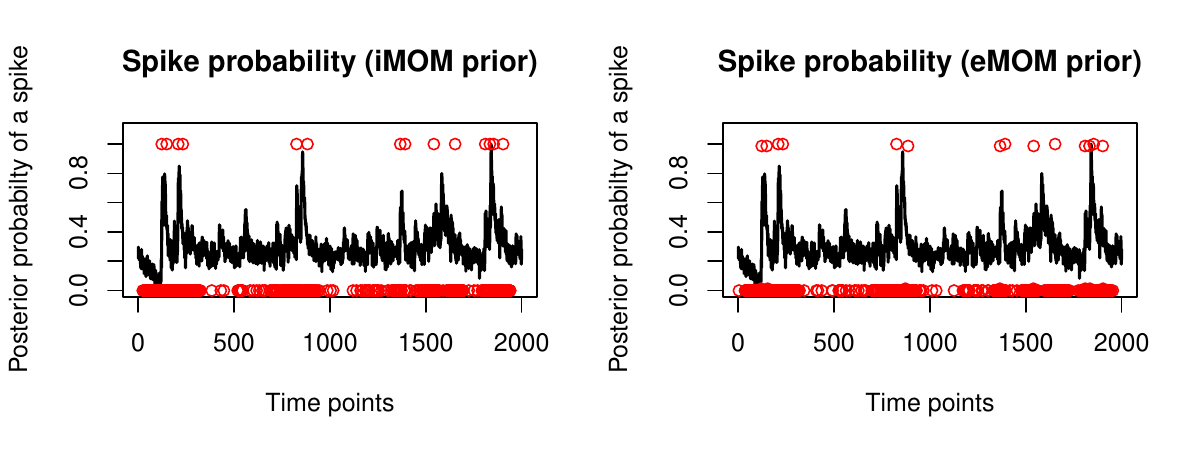} }}%
    \caption{\textbf{Mice data:} \emph{The calcium fluorescence data from mice in \cite{Theis}. The two panel displays the centered and scaled DF/F-transformed fluorescence re scaled to $(0,1)$ (in black lines) and the posterior probabilities of  spikes at different time points by red dots, for the iMOM, eMOM based methods respectively.}\label{data1_UQ}}
\end{figure}

\subsection{Application to \cite{Theis} data}
\begin{figure}[!htb]
    \centering
    {{\includegraphics[width=16cm, height = 10cm]{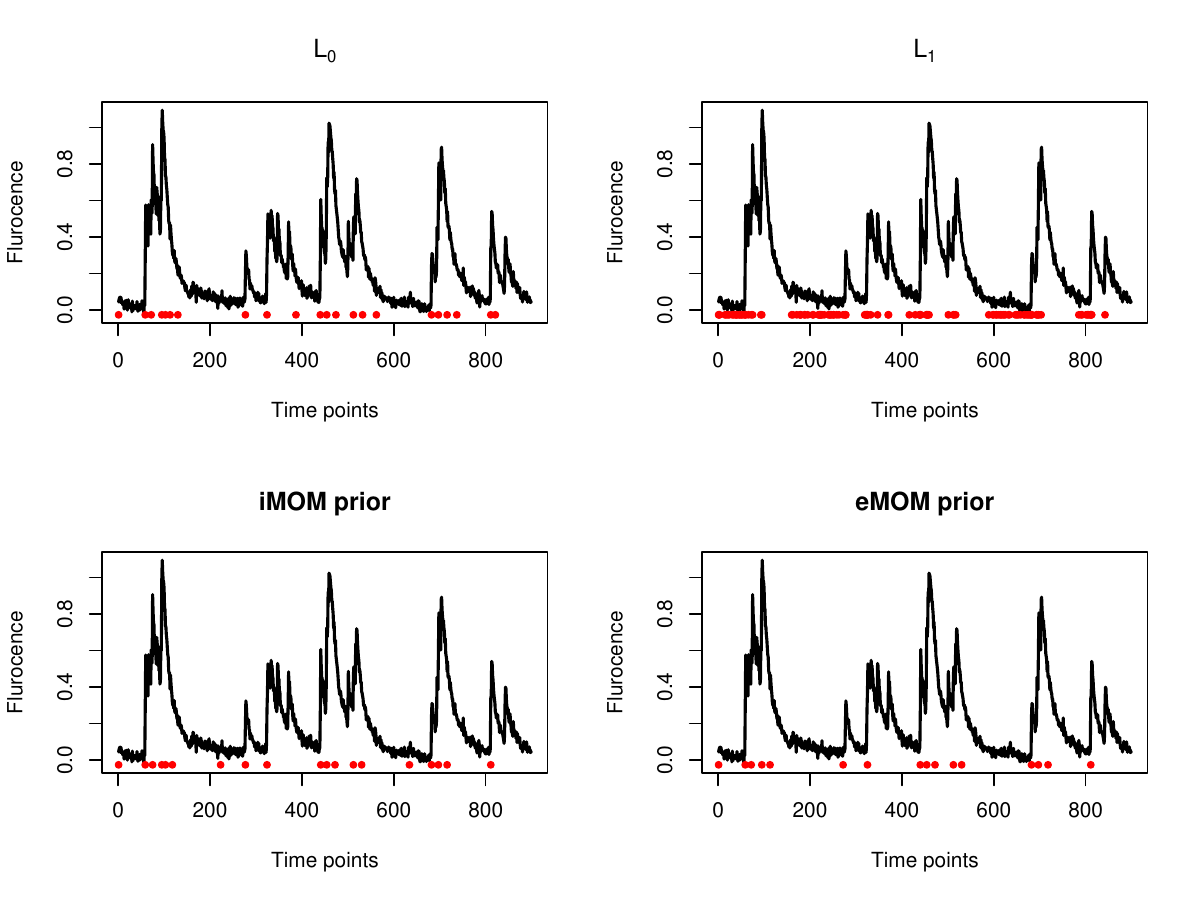} }}%
    \caption{\textbf{Zebra fish data:} \emph{The calcium fluorescence data for Zebra fish in \cite{Rupprecht}. Each panel displays the DF/F-transformed fluorescence (in black line ), the estimated spikes by the red dots. The four panels display results obtained via the $L_0$, $L_1$, iMOM, eMOM based methods respectively. }\label{data4_estimation}}
\end{figure}

First, we consider a data set that consists of simultaneous calcium imaging of \emph{mice}, originally presented in \cite{Theis}, and compiled in \cite{Rupprecht} as a part of the large and diverse database. The black lines in the figure \ref{data4_estimation} presents the first 28 minutes of calcium trace recording from  cell 10 of mice 2 for, which expresses OGB-1. The data are measured  for a total of 2000 time-steps. As earlier, the raw fluorescence traces are DF/F transformed with a 20\% percentile filter (\cite{Friedrich2017}).

For this data set, the $L_0$, $L_1$, inverse moment prior, and exponential moment prior based approaches identified $15$, $38$, $14$ and $14$ spikes  respectively, presented as red dots in  figure \ref{data1_estimation}. The spikes corresponding to the  the $L_0$, inverse moment prior, and exponential moment prior based approaches appear at similar and visually reasonable time points. But, as observed in our simulation studies as well as much of previous literature (\cite{Jewell2018}), the $L_1$ regularised approach results in many, typically small in magnitude, false spike discoveries. Unlike the penalised likelihood based procedures, our proposed methods provide complete uncertainty quantification regarding the spike train inference. In particular, the reported maximum a posteriori estimates corresponding to the inverse moment and exponential moment priors has posterior probabilities equal to $0.986$ and $0.999$ respectively. Further, the posterior probability of a spike at the different time point corresponding to the inverse moment and exponential moment prior based approaches are presented as red dotsin the two panels in figure \ref{data1_UQ}.

\begin{figure}[!htb]
    \centering
    {{\includegraphics[width=16cm, height = 6cm]{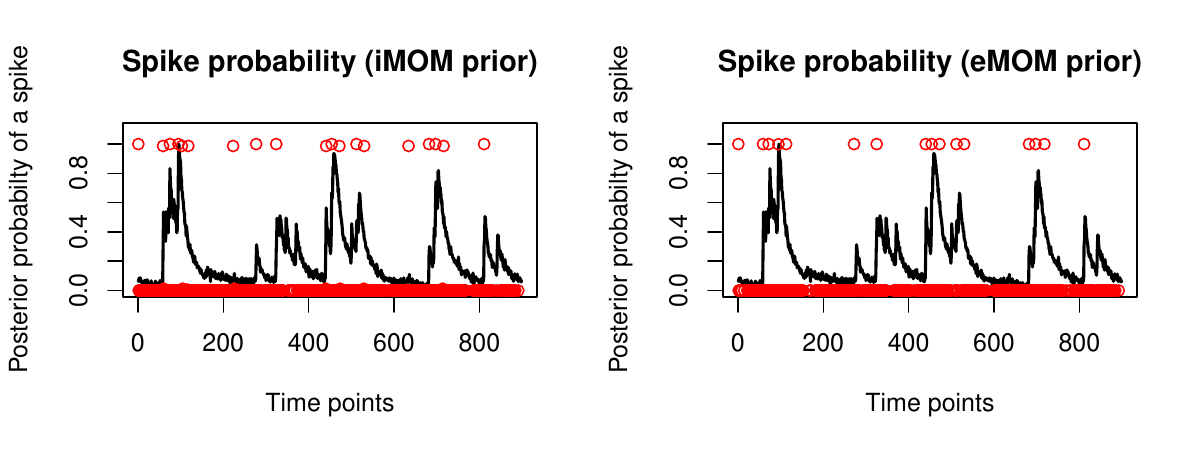} }}%
    \caption{\textbf{Zebra fish data:} \emph{The calcium fluorescence data for Zebra fish in \cite{Rupprecht}. The two panel displays the centered and scaled DF/F-transformed fluorescence re scaled to $(0,1)$ (in black lines) and the posterior probabilities of  spikes at different time points by red dots, for the iMOM, eMOM based methods respectively.}\label{data4_UQ}}
\end{figure}

\subsection{Application to \cite{Rupprecht} data}
Next, we consider a data set that consists of simultaneous calcium imaging of \emph{zebrafish}, originally presented in \cite{Rupprecht}, as a part of a large and diverse database. The black lines in the figure \ref{data4_estimation} presents a 81 minute calcium trace recording from  cell 4 of fish 2 for, which expresses OGB-1. The data are measured  for a total of 900 time-steps. The raw fluorescence traces are DF/F transformed with a 20\% percentile filter (\cite{Friedrich2017}).

For this data set, the $L_0$, $L_1$, inverse moment prior, and exponential moment prior based approaches identified $22$, $120$, $19$ and $16$ spikes  respectively, presented as red dots in  figure \ref{data4_estimation}. The spikes corresponding to the  the $L_0$, inverse moment prior, and exponential moment prior based approaches appear at similar and visually reasonable time points. But, as observed in our simulation studies as well as much of previous literature (\cite{Jewell2018}), the $L_1$ regularised approach results in many, typically small in magnitude, false spike discoveries. Unlike the penalised likelihood based procedures, our proposed methods provide complete uncertainty quantification regarding the spike train inference. In particular, the reported maximum a posteriori estimates corresponding to the inverse moment and exponential moment priors has posterior probabilities equal to $0.987$ and $0.998$ respectively. Further, the posterior probability of a spike at the different time point corresponding to the inverse moment and exponential moment prior based approaches are presented as red dotsin the two panels in figure \ref{data4_UQ}.

%\clearpage
\section{Discussion}
Recent advances in neuroscience have enabled researchers to measure the activities of large numbers of neurons simultaneously in behaving animals. Scientists have access to the fluorescence of each of the neurons that provides a first-order approximation to the neural activity over time. Determining a neuron's exact spike time is an important challenge that remains an active area of research within the field of computational neuroscience. This problem has recently been addressed using penalized likelihood methods (\cite{Jewell2018}, \cite{NIPS2016_fc2c7c47}, \cite{Friedrich2017}. In this paper, we proposed a new approach for this task based on a novel application of  a mixture prior on spike variables that has  a point mass at zero and a half non local density (\cite{Johnson2010}), in the popular auto-regressive model for calcium dynamics. Our proposals lead to substantial improvements over the previous proposal based on $L_{1}$ regularization , and enjoys comparable estimation accuracy to the state-of-the-art $L_{0}$ proposal,  in simulations as well as on recent calcium imaging data sets. Notably, contrary to most optimization based frequentist approaches, our methodology yields automatic uncertainty quantification associated with the spike train inference.

One compelling direction for future enquiry includes positing a hierarchical Bayesian framework to simultaneously model multitude of  spike trains arising from different neurons an animal interest. This will lead us to understand the complete neural response of the animal towards an external signal. Further, there is merit in attempting to carry out the spike train inference in online fashion, i.e, real time estimation and uncertainty quantification of spikes as the data arrives to a system, in order to improve its practical applicability.

%\section*{Appendix}
% Choice of \gamma
%\clearpage
\section{Data availability statement}
All the data utilized in this study is freely available in:\\
 \href{https://github.com/HelmchenLabSoftware/Cascade/}{github.com/HelmchenLabSoftware/Cascade}.
\section*{Disclosure statement}
The author reports there are no competing interests to declare.
\section*{Funding details}
There was non external on internal funding for this work.

\bibliography{paper-ref}
\bibliographystyle{apalike}

%\clearpage

\end{document}